\documentclass[12pt]{article}
\usepackage{epsf}
\title{ {\bf Lepton flavor violating
$l_i\rightarrow l_j\,\gamma\gamma$ decays induced by scalar
unparticle}}
\author{\vspace{1cm}\\
        {\bf E. O. Iltan,}
        \thanks{E-mail address:
        eiltan@newton.physics.metu.edu.tr}
 \\
        Physics Department, Middle East Technical University \\
        Ankara, Turkey\\}

\date{}

\begin{document}
\setlength{\baselineskip}{24pt}
\maketitle
\setlength{\baselineskip}{7mm}
\begin{abstract}
We study the radiative lepton flavor violating  $l_i\rightarrow
l_j\,\gamma\gamma$ decays in the case that the lepton flavor
violation is induced by the scalar unparticle mediation. We
restrict the scaling dimension $d_u$ and the scalar
unparticle-photon-photon coupling by using the experimental upper
limit of the branching ratio of the decay $\mu\rightarrow
e\,\gamma\gamma$. Furthermore, we predict the BRs of the other
radiative decays by using the restrictions we get. We observe that
the measurements of upper limits of BRs of these decays  ensure
considerable information for testing the possible signals coming
from unparticle physics.
\end{abstract}
\thispagestyle{empty}
\newpage
\setcounter{page}{1}
%
The radiative decays with two photon output are interesting to
analyze since they are rich due to two different polarizations of
photons, namely the parallel and the perpendicular ones. The
measurement of the photon spin polarization provides comprehensive
information about the free parameters existing in the model used.
In the present work we study the radiative lepton decays with two
photon output, i.e, $l_i\rightarrow l_j\,\gamma\gamma$, $i\neq j$.
These processes are driven by a lepton flavor violating (LFV)
mechanism and, in the framework of the SM, the lepton mixing
matrix which arises with non-zero neutrino masses is responsible
for this violation. In this case these decays exist at least in
the loop level and their branching ratios (BRs) are highly
suppressed. Here, for the lepton flavor violation, we consider the
another mechanism based on the unparticle physics.

The unparticle stuff is proposed by Georgi \cite{Georgi1,Georgi2}
and the so called unparticle, which looks like a number of $d_u$
massless invisible particles, has non-integer scaling dimension
$d_u$. The unparticle stuff is the low energy manifestation of a
hypothetical non-trivial scale invariant ultraviolet sector,
having a non-trivial infrared fixed point. The interactions of
unparticles with the SM fields are driven  by the effective
lagrangian in the low energy level and the corresponding
Lagrangian reads
\begin{equation}
{\cal{L}}_{eff}\sim
\frac{\eta}{\Lambda_U^{d_u+d_{SM}-n}}\,O_{SM}\, O_{U} \,,
\label{efflag}
\end{equation}
where $O_U$ is the unparticle operator, the parameter $\eta$ is
related to the energy scale of ultraviolet sector, the low energy
one and the matching coefficient \cite{Georgi1,Georgi2,Zwicky} and
$n$ is the space-time dimension.

The search for unparticle(s) ensures a valuable information about
the expected ultraviolet sector and the scale invariance. The
missing energies at various processes which can be measured at LHC
or $e^+e^-$ colliders, the dipole moments of fundamental particles
are among the possible candidates for searching the effects of
unparticle(s). In the literature there is an extensive
phenomenological work done on unparticles
\cite{Georgi2}-\cite{XZang}. These studies are about the possible
effects of unparticle stuff on the missing energy of many
processes, the anomalous magnetic moments, the electric dipole
moments, $D^0-\bar{D}^0$ and $B^0-\bar{B}^0$ mixing, lepton flavor
violating interactions, direct CP violation in particle physics,
cosmology and astrophysics.

In the present work, we consider that the LF violation is switched
on with the scalar unparticle ($\textit{U}$)-lepton-lepton vertex.
Furthermore, we expect that $\textit{U}$-photon ($\gamma$)-photon
($\gamma$) interaction exists and, finally, the radiative
$l_i\rightarrow l_j\,\gamma\gamma$, $i\neq j$ decays appear in the
tree level, with the scalar unparticle mediation. In our
calculations, we respect the experimental upper limit of the BR of
the $\mu\rightarrow e\,\gamma\gamma$ decay, BR$< 7.2\,10^{-11}\,
90\% \,CL$ \cite{partdatagroup} and try to restrict the scaling
dimension $d_u$ and the $\textit{U}-\gamma-\gamma$ coupling. In
addition to this, we predict the BRs of the other radiative decays
by using the restrictions we get.

Now, we start by choosing the appropriate operators with the
lowest possible dimension since they have the most powerful effect
in the low energy effective theory (see for example \cite{SChen}).
The effective interaction lagrangian driving the LFV decays reads
\begin{eqnarray}
{\cal{L}}_1= \frac{1}{\Lambda_U^{du-1}}\Big (\lambda_{ij}^{S}\,
\bar{l}_{i} \,l_{j}+\lambda_{ij}^{P}\,\bar{l}_{i}
\,i\gamma_5\,l_{j}\Big)\, O_{U} \, , \label{lagrangianscalar}
\end{eqnarray}
where $l$ is the lepton field,  $O_{U}$ is the scalar unparticle
(\textit{U}) operator and $\lambda_{ij}^{S}$ ($\lambda_{ij}^{P}$)
is the scalar (pseudoscalar)  coupling. On the other hand, the
effective Lagrangian which is responsible for two photon radiation
is
\begin{eqnarray}
{\cal{L}}_2= \frac{1}{\Lambda_U^{du}}\Big (\lambda_0\,
F_{\mu\nu}\,F^{\mu\nu}+\lambda_0^{\prime}\,
\tilde{F}_{\mu\nu}\,F^{\mu\nu} \Big)\, O_{U} \, ,
\label{lagrangianphoton}
\end{eqnarray}
where $F_{\mu\nu}$ is the electromagnetic field tensor and
$\tilde{F}_{\mu\nu}=\frac{1}{2}\epsilon_{\mu\nu\alpha\beta}\,
F^{\alpha\beta}$. In radiative two photon decays the outgoing
photons are in one of the possible states given by
$F^{\mu\nu}\,F_{\mu\nu}$ and $\tilde{F}_{\mu\nu}\,F^{\mu\nu}$ and
these states corresponds to the parallel ($\mathbf{\epsilon}_1 .
\mathbf{\epsilon}_2$) and perpendicular
($\mathbf{\epsilon}_1\times \mathbf{\epsilon}_2$) spin
polarizations of photons which are regulated by the couplings
$\lambda_0$ and $\lambda_0^{\prime}$ in the present case. In our
calculations we consider a parameter $\alpha$ such that
$\lambda_0^{\prime}= \alpha \,\lambda_0$.

The tree level $l_i\rightarrow l_j \gamma\gamma $ decay (see Fig.
1) is carried by connecting the LFV vertex\footnote{The vertex
factor:
$\frac{i}{\Lambda_U^{d_u-1}}(\lambda^S+i\,\gamma_5\,\lambda^P)$.}
and the $\textit{U}-\gamma-\gamma$ vertex\footnote{The vertex
factor: $\frac{4\,i}{\Lambda_U^{d_u}}\,\Bigg(\lambda_0\,
(k_{1\nu}\,k_{2\mu}-k_1.k_2\,g_{\mu\,\nu})+
\lambda_0^{\prime}\,\epsilon_{\alpha\beta\mu\nu}\,k_1^\alpha\,k_2^\beta
\Bigg)$ where $k_{1(2)}$ is the four momentum of photon with
polarization vector $\epsilon_{1\,\mu \,(2\,\nu)}$.} by the scalar
unparticle propagator which is obtained by using the scale
invariance. The two point function of the unparticle reads
\cite{Georgi2, Cheung1}
\begin{eqnarray}
\int\,d^4x\,
e^{ipx}\,<0|T\Big(O_U(x)\,O_U(0)\Big)0>=i\frac{A_{d_u}}{2\,\pi}\,
\int_0^{\infty}\,ds\,\frac{s^{d_u-2}}{p^2-s+i\epsilon}=i\,\frac{A_{d_u}}
{2\,sin\,(d_u\pi)}\,(-p^2-i\epsilon)^{d_u-2} \, ,
\label{propagator}
\end{eqnarray}
with the factor $A_{d_u}$
\begin{eqnarray}
A_{d_u}=\frac{16\,\pi^{5/2}}{(2\,\pi)^{2\,d_u}}\,\frac{\Gamma
(d_u+\frac{1}{2})} {\Gamma(d_u-1)\,\Gamma(2\,d_u)} \, .
\label{Adu}
\end{eqnarray}
The function $\frac{1}{(-p^2-i\epsilon)^{2-d_u}}$ in eq.
(\ref{propagator}) becomes
\begin{eqnarray}
\frac{1}{(-p^2-i\epsilon)^{2-d_u}}\rightarrow
\frac{e^{-i\,d_u\,\pi}}{(p^2)^{2-d_u}} \, , \label{strongphase}
\end{eqnarray}
for $p^2>0$ and a non-trivial phase appears as a result of
non-integral scaling dimension.

Now, we present the decay amplitude for the radiative decay
$l_i\rightarrow l_j\gamma\gamma$:
\begin{eqnarray}
M(l_i\rightarrow
l_j\gamma\gamma)=\epsilon_1^{\mu}(k_1)\,\epsilon_2^{\nu}(k_2)\,\bar{l}_j(p')\,
T_{\mu\nu}\,l_i(p) \, ,\label{Ampl1}
\end{eqnarray}
where the structure $T_{\mu\nu}$ reads
\begin{eqnarray}
T_{\mu\nu}=\frac{-A_{d_u}\,(-q^2)^{d_u-2}}{2\,sin\,(d_u\,\pi)\,
\lambda_U^{2d_u-1}}\,\Big(\lambda_{ij}^{S}+\gamma_5\,\lambda_{ij}^{P}\Big)\,
A_{\mu\nu}\, , \label{Tmunu}
\end{eqnarray}
with
\begin{eqnarray}
A_{\mu\nu}=4\,\Big( (k_{1\nu}\,k_{2\mu}-k_1.k_2\,
g_{\mu\nu})\,\lambda_0 + \epsilon_{\alpha\beta\mu\nu}\,
k_1^{\alpha}\,k_2^{\beta}\,\lambda_0^{\prime} \Big) \, .
\label{Amunu}
\end{eqnarray}
Finally, the partial decay with $d\Gamma$ for $l_i\rightarrow l_j
\,\gamma\gamma$ decay can be obtained by using the matrix element
square $|M|^2$ as
\begin{eqnarray}
d\Gamma=\frac{1}{128\,\pi^3\,m_{l_i}}\,|M|^2\, dE_1\,dE_j\, ,
\label{dGamma}
\end{eqnarray}
where $E_1 (E_j)$ is the energy of the photon with polarization
four vector $\epsilon_1^{\mu}(k_1)$ (the outgoing lepton). In our
numerical calculations we analyze the BRs of the decays under
consideration by using the total decay widths of incoming leptons
\cite{partdatagroup} (see Table I).
%
\\ \\
{\Large \textbf{Discussion}}
\\ \\
This section is devoted to  analysis of the radiative LFV
$l_i\rightarrow l_j\,\gamma\gamma$ decays. Here, we consider that
the LF violation is switched on with the
$\textit{U}$-lepton-lepton coupling, in the framework of the
effective theory. On the other hand, the possible
$\textit{U}-\gamma-\gamma$ vertex results in that the decays under
consideration exist even in the tree level, with the unparticle
mediation. Therefore, the physical quantities like the BRs of
these decays can be informative in the determination of the free
parameters, the scaling dimension of the unparticle, the couplings
and the energy scale in the scenario studied. Here, we choose the
scaling dimension $d_u$ in the range\footnote{$d_u>1$ is chosen
since one is free from the non-integrable singularity problem in
the decay rate \cite{Georgi2}. Furthermore, the momentum integrals
converge for $d_u<2$ \cite{Liao1}} $1< d_u <2$. For off diagonal
$\textit{U}$-lepton-lepton couplings $\lambda_{ij}^S$ and
$\lambda_{ij}^P$\footnote{In the following we drop the indices $i$
and $j$.} we take the numerical values of the order of
$10^{-3}-10^{-1}$ and we predict the $\textit{U}-\gamma-\gamma$
coupling $\lambda_0$ by restricting the BR($\mu\rightarrow e
\gamma\gamma$) to its current experimental upper limit, by taking
the energy scale at the order of $\Lambda_U=10\,(TeV)$. This
analysis restricts the pair of parameters, the scaling dimension
$d_u$ and the coupling $\lambda_0$. Furthermore, we estimate the
BRs of the LFV radiative decays $\tau\rightarrow e \gamma\gamma$
and $\tau\rightarrow \mu \gamma\gamma$, by using the restriction
obtained for the pair $d_u$ and $\lambda_0$.
Notice that throughout our calculations we use the input values
given in Table (\ref{input}).
\begin{table}[h]
        \begin{center}
        \begin{tabular}{|l|l|}
        \hline
        \multicolumn{1}{|c|}{Parameter} &
                \multicolumn{1}{|c|}{Value}     \\
        \hline \hline
        $m_e$           & $0.0005$   (GeV)  \\
        $m_{\mu}$                   & $0.106$ (GeV) \\
        $m_{\tau}$                  & $1.780$ (GeV) \\
        $\Gamma^{Tot}_\mu$           & $2.99\times 10^{-19}$ (GeV) \\
        $\Gamma^{Tot}_\tau$           & $2.26\times 10^{-10}$ (GeV) \\
        \hline
        \end{tabular}
        \end{center}
\caption{The values of the input parameters used in the numerical
          calculations.}
\label{input}
\end{table}
\newpage
In  Fig.\ref{muegamgamlam0du}, we present the magnitude of the
coupling $\lambda_0$ with respect to $d_u$ for the fixed value BR
$(\mu\rightarrow e \gamma\,\gamma)=7.2\times 10^{-11}$ and the
energy scale $\Lambda_U=10\, (TeV)$. Here the solid-dashed-small
dashed (upper dotted-intermediate dotted-lower dotted) line
represents $\lambda_0$ for $\lambda^S=\lambda^P=0.001-0.01-0.1$
and $\alpha=1$ ($\alpha=0$).  The coupling $\lambda_0$, which is
sufficient to get the experimental upper limit of the BR, becomes
stronger for the larger values of $d_u$. This is due to the fact
that the BR is strongly sensitive to the scaling dimension $d_u$
and it is suppressed with its increasing values. For
$\lambda^S=\lambda^P=0.001$ the coupling $\lambda_0$ should be in
the range $0.001-1.0$ for the scaling dimension $d_u$, $1.1< d_u <
1.4$. In the case of strong couplings $\lambda^S=\lambda^P=0.01\,
(0.1)$ this range is obtained for larger scaling dimension, $d_u <
1.5\,(1.6)$. Furthermore, it is observed that the amount of
coupling $\lambda_0$ is weakly sensitive to the parameter $\alpha$
for its values that is less than one. Fig.\ref{muegamgamalfdu}
represents the magnitude of the parameter $\alpha^2$ with respect
to $d_u$ for the fixed value BR $(\mu\rightarrow e
\gamma\,\gamma)=7.2\times 10^{-11}$ and the energy scale
$\Lambda_U=10\, (TeV)$. Here the solid (dashed, small dashed) line
represents $\alpha^2$ for $\lambda^S=\lambda^P=0.001\, (0.01,
0.1)$ and $\lambda_0=0.01$. For the selected numerical value of
the coupling $\lambda_0$, $\lambda_0=0.01$, the parameter $\alpha$
should be greater than one for $\lambda^S=\lambda^P=0.001$ and
$0.01$. This is the case that the perpendicular spin polarization
exceeds the parallel spin polarization for two photon system.
Furthermore, the scaling parameter $d_u$ should be $>1.18$ and
$>1.27$ in order to get a solution. For $\lambda^S=\lambda^P=0.1$,
$\alpha$ can be less than one where the parallel spin polarization
exceeds the perpendicular spin polarization for two photon system.
This is the case that $d_u$ should be $>1.34$. These observations
are interesting since the more accurate forthcoming measurement of
the BR of the decay under consideration would ensure valuable
information about the possible $\textit{U}-\gamma-\gamma$ coupling
and the scaling dimension $d_u$. In addition to this, the precise
determination of the photon polarization in the experiments would
be informative in the determination of these parameters.

Fig.\ref{taumutauegamgamduLogy2} is devoted to the BRs of the
decays $\tau\rightarrow e \gamma\,\gamma$ and $\tau\rightarrow \mu
\gamma\,\gamma$ with respect to $d_u$ for
$\lambda^S=\lambda^P=0.01$, $\alpha=1$ and $\Lambda_U=10\, (TeV)$.
Here the solid (dashed) line represents the BR for the pair $d_u$
and $\lambda_0$ which is obtained by using the restriction  BR
$(\mu\rightarrow e \gamma\,\gamma)=7.2\times 10^{-11}$
$(10^{-12})$. The BRs for both decays almost coincides and they
are of the order of $10^{-13}$\, ($10^{-15}$) for the pair
$\lambda_0\sim 1.0$-$d_u\sim 1.5$. These numerical values are
extremely small, however, there is a chance to increase them by
considering that the off diagonal couplings of scalar
$\textit{U}$-lepton-lepton couplings are not flavor blind and
sensitive to the lepton flavor.

For completeness, in  Fig.\ref{tauegamgamdu} we plot the BRs of
$\tau\rightarrow e \gamma\,\gamma$ and $\tau\rightarrow \mu
\gamma\,\gamma$ decays with respect to $d_u$ for
$\lambda^S=\lambda^P=0.01$, $\alpha=1$ in the case that there is
no restriction for the pair $d_u$ and $\lambda_0$. Here the solid
(dashed, small dashed) line represents the BRs for
$\lambda_0=0.01$ and $\Lambda_U=1.0\, (5.0,\, 10)\, (TeV)$. It is
observed that the BRs of both decays are almost coincides and they
are in the range $10^{-12}-10^{-6}$ ($10^{-15}-10^{-8}$,
$10^{-16}-10^{-9}$) for the energy scale $\Lambda_U=1.0\, (5.0,\,
10)\, (TeV)$ and the interval $1.1<d_u<1.5$.

As a summary, the radiative LFV decays $l_i\rightarrow
l_j\gamma\gamma$ can exist in the tree level with the help of the
unparticle idea. The measurements of upper limits of BRs of these
decays (the more accurate one for the $\mu\rightarrow e
\gamma\,\gamma$ decay) would be instructive for testing the
possible signals coming from the new physics which drives the
flavor violation and they would ensure considerable information on
the restriction of free parameters.
\newpage
\begin{figure}[htb]
\vskip 6.2truein\centering \epsfxsize=4.5in
\leavevmode\epsffile{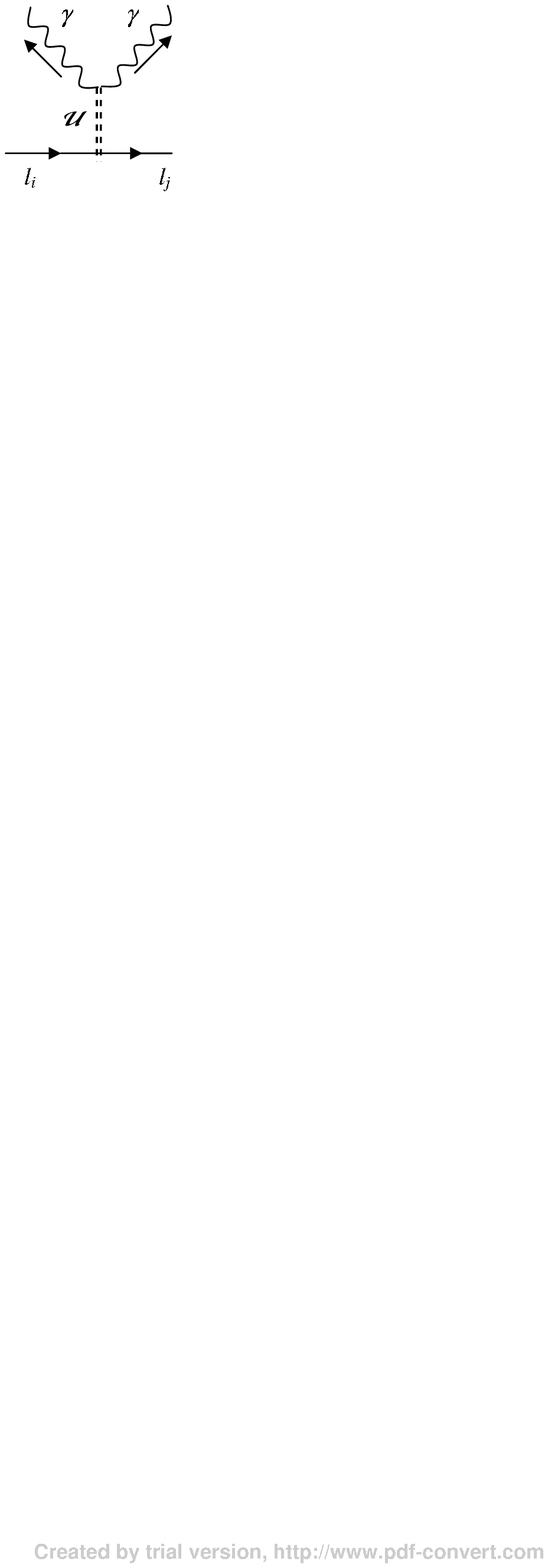} \vskip -11.5truein
\caption[]{Tree level diagram contribute to $l_i\rightarrow l_j
\gamma\gamma$ decay with scalar unparticle mediator. Solid line
represents the lepton field, wavy line the photon field, double
dashed line the scalar unparticle field.} \label{l1l2gamgam}
\end{figure}
\newpage
\begin{figure}[htb]
\vskip -3.0truein \centering \epsfxsize=6.8in
\leavevmode\epsffile{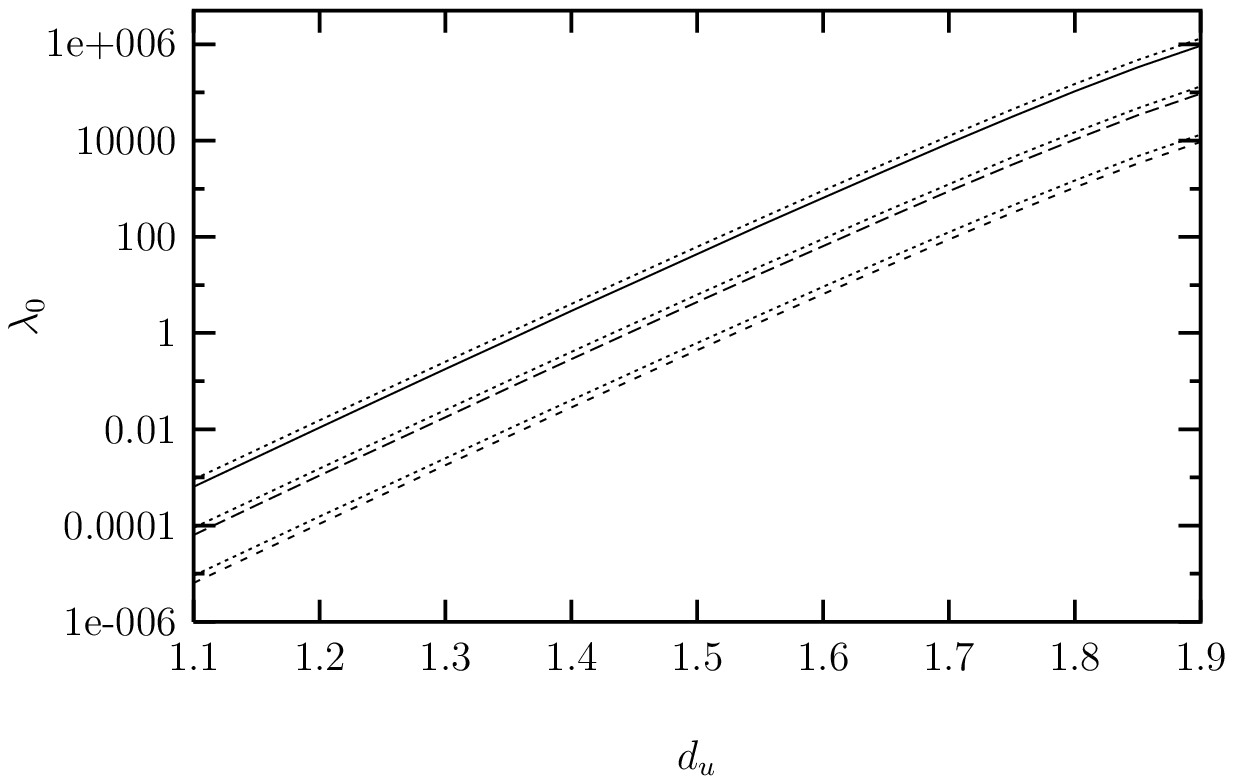} \vskip -3.0truein
\caption[]{$\lambda_0$ with respect to $d_u$ for the fixed value
BR $(\mu\rightarrow e \gamma\,\gamma)=7.2\times 10^{-11}$ and the
energy scale $\Lambda_U=10\, (TeV)$. Here the solid-dashed-small
dashed (upper dotted-intermediate dotted-lower dotted) line
represents $\lambda_0$ for $\lambda^S=\lambda^P=0.001-0.01-0.1$
and $\alpha=1$ ($\alpha=0$).} \label{muegamgamlam0du}
\end{figure}
\begin{figure}[htb]
\vskip -3.0truein \centering \epsfxsize=6.8in
\leavevmode\epsffile{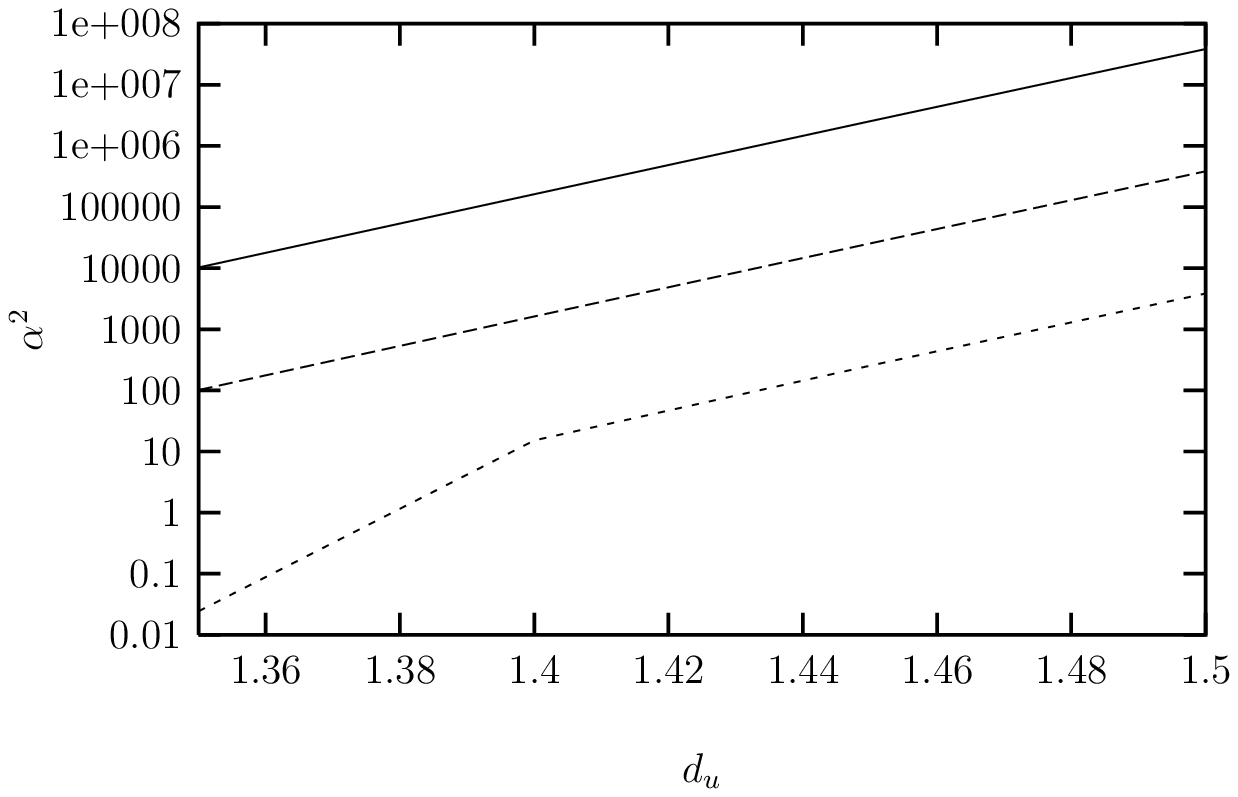} \vskip -3.0truein
\caption[]{$\alpha^2$ with respect to $d_u$ for the fixed value BR
$(\mu\rightarrow e \gamma\,\gamma)=7.2\times 10^{-11}$ and the
energy scale $\Lambda_U=10\, (TeV)$. Here the solid (dashed, small
dashed) line represents $\alpha^2$ for
$\lambda^S=\lambda^P=0.001\, (0.01, 0.1)$ and $\lambda_0=0.01$.}
\label{muegamgamalfdu}
\end{figure}
\begin{figure}[htb]
\vskip -3.0truein \centering \epsfxsize=6.8in
\leavevmode\epsffile{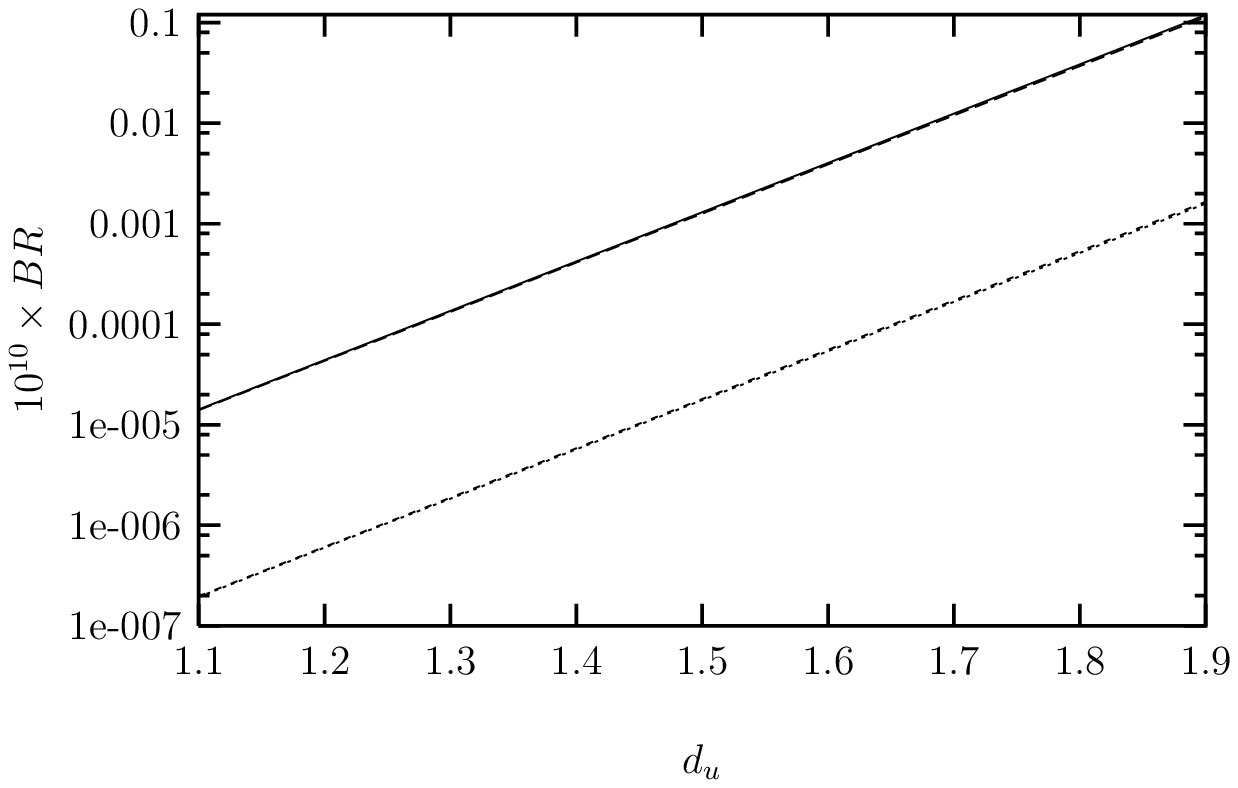} \vskip -3.0truein
\caption[]{BR$(\tau\rightarrow e (\mu) \gamma\,\gamma)$  with
respect to $d_u$ for $\lambda^S=\lambda^P=0.01$, $\alpha=1$ and
$\Lambda_U=10\, (TeV)$. Here the solid (dashed) line represents
the BR for the pair $d_u$ and $\lambda_0$ which is obtained by
using the restriction  BR $(\mu\rightarrow e
\gamma\,\gamma)=7.2\times 10^{-11}$ $(10^{-12})$.}
\label{taumutauegamgamduLogy2}
\end{figure}
\begin{figure}[htb]
\vskip -3.0truein \centering \epsfxsize=6.8in
\leavevmode\epsffile{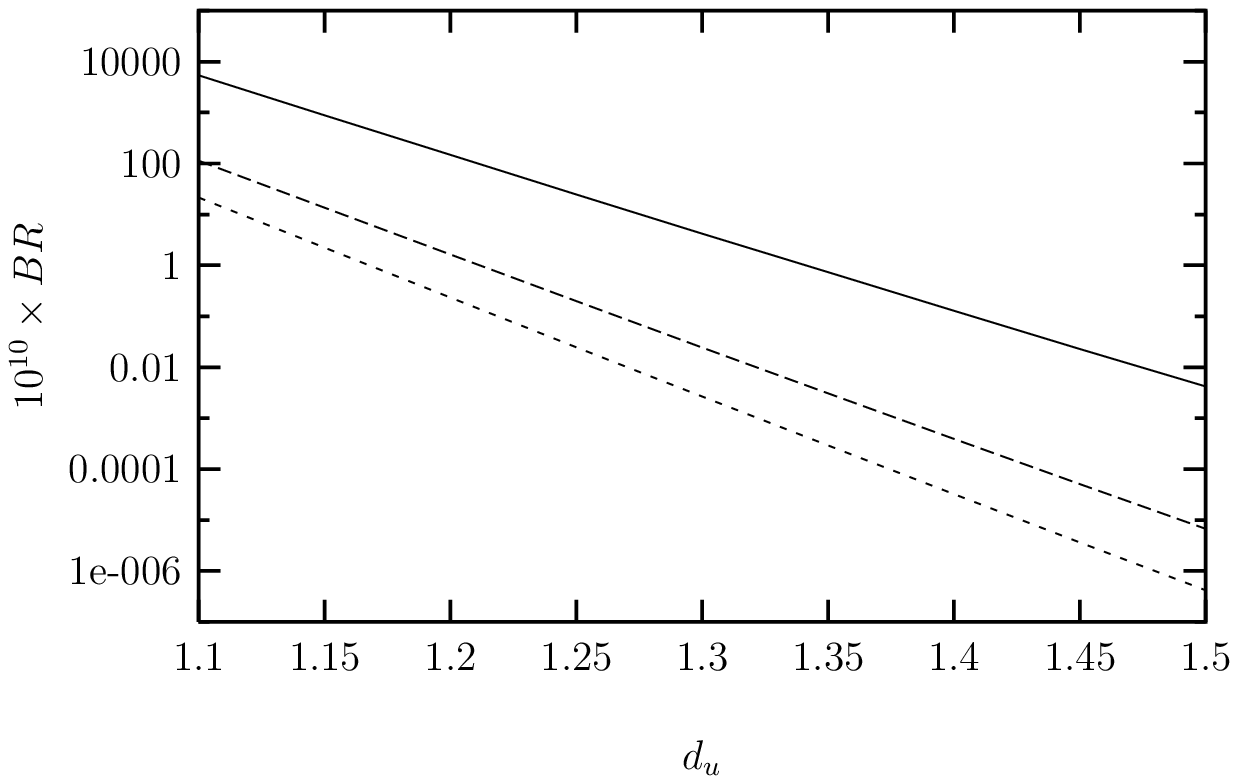} \vskip -3.0truein
\caption[]{The BR $(\tau\rightarrow e (\mu) \gamma\,\gamma)$ with
respect to $d_u$ for $\lambda^S=\lambda^P=0.01$, $\alpha=1$. Here
the solid (dashed, small dashed) line represents the BR for
$\lambda_0=0.01$ and $\Lambda_U=1.0\, (5.0,\, 10)\, (TeV)$.}
\label{tauegamgamdu}
\end{figure}
\end{document}